\documentclass{ws-procs9x6}

\begin{document}

\title{Algebraic aspects of the correlation functions 
of the integrable higher-spin XXZ spin chains with arbitrary entries}

\author{Tetsuo Deguchi$^{*}$}

\address{Department of Physics, 
Graduate School of Humanities and Sciences, Ochanomizu University,\\
2-1-1 Ohtsuka, Bunkyo-ku, Tokyo 112-8610, Japan \\
$^{*}$E-mail: deguchi@phys.ocha.ac.jp}

\author{Chihiro Matsui $^{\dagger \, \, \ddagger}$} 
\address{Department of Physics, Graduate School of Science, the University of Tokyo \\ 7-3-1 Hongo, Bunkyo-ku, Tokyo 113-0033, Japan \\ 
$^{\dagger}$ CREST, JST, 4-1-8 Honcho Kawaguchi, Saitama, 332-0012, Japan \\
$^{\ddagger}$E-mail: matsui@spin.phys.s.u-tokyo.ac.jp}

\begin{abstract} 
We discuss some fundamental properties of the XXZ spin chain,   
which are important in the algebraic Bethe-ansatz derivation for  
the multiple-integral representations of the spin-$s$ XXZ correlation 
function with an arbitrary product of elementary matrices \cite{DM2}.  
For instance, we construct Hermitian conjugate vectors 
in the massless regime 
and introduce the spin-$s$ Hermitian elementary matrices. 
\end{abstract}


\bodymatter

%
%
\setcounter{section}{0} 
\setcounter{equation}{0} 
 \renewcommand{\theequation}{1.\arabic{equation}}
\section{Introduction}

The correlation functions of the spin-1/2 XXZ spin chain have 
attracted much interest in mathematical physics 
through the last two decades. 
One of the most fundamental results is the exact derivation of 
their multiple-integral representations.  
The multiple-integral representations of the XXZ correlation functions 
were derived for the first time by making use of the $q$-vertex operators 
through the affine quantum-group symmetry 
in the massive regime for the infinite lattice at zero temperature 
\cite{Miki,Jimbo-Miwa}. 
They were also derived in the massless regime 
by solving the $q$-KZ equations \cite{Jimbo-Miwa-qKZ,Takeyama}. 
Making use of algebraic Bethe-ansatz  techniques such as scalar products 
\cite{Slavnov,Korepin,MS2000,KMT1999,MT2000},   
the multiple-integral representations were 
derived for the XXZ correlation functions 
under a non-zero magnetic field. \cite{KMT2000} 
They were extended into those 
at finite temperatures \cite{Goehmann-CF}, 
and even for a large finite chain \cite{Damerau}. 
Interestingly, they are factorized 
in terms of single integrals. \cite{Jimbo-Miwa-Smirnov} 
Furthermore, the asymptotic expansion of a correlation function 
of the XXZ model has been systematically discussed \cite{KMTK2008}. 
Thus, the exact study of the XXZ correlation functions 
should play an important role not only 
in the mathematical physics of integrable models 
but also in many areas of theoretical physics.

The Hamiltonian of the spin-1/2 XXZ spin chain 
under the periodic boundary conditions is given by  
\begin{equation} 
{\cal H}_{\rm XXZ} =  
{\frac 1 2} \sum_{j=1}^{L} \left(\sigma_j^X \sigma_{j+1}^X +
 \sigma_j^Y \sigma_{j+1}^Y + \Delta \sigma_j^Z \sigma_{j+1}^Z  \right) \, . 
\label{hxxz}
\end{equation}
Here $\sigma_j^{a}$ ($a=X, Y, Z$) are the Pauli matrices defined 
on the $j$th site and $\Delta$ denotes the XXZ coupling.   
We define  parameter $q$  by  
\begin{equation} 
\Delta= (q+q^{-1})/2 \, . 
\end{equation}
We define $\eta$ by $q=\exp \eta$. 
In the massive regime: $\Delta > 1$, we put $\eta=\zeta$ with 
$\zeta > 0$. 
At $\Delta=1$ (i.e. $q=1$)  the Hamiltonian (\ref{hxxz}) gives 
 the antiferromagnetic Heisenberg (XXX) chain. 
In the massless regime: $-1 < \Delta \le 1$, 
we set $\eta= i \zeta$,  and we have $\Delta= \cos \zeta$ with 
$0 \le  \zeta < \pi$ for the spin-1/2 XXZ spin chain (\ref{hxxz}).  
In the paper we consider a massless region: $0 \le \zeta < \pi/2s$ for 
the ground-state of the integrable spin-$s$ XXZ spin chain.

Recently, the correlation functions and form factors of the 
integrable higher-spin XXX and XXZ spin chains 
have been derived by the algebraic Bethe-ansatz method 
\cite{Kitanine2001,Castro-Alvaredo,DM1,DM2}.  
The solvable higher-spin generalizations of the XXX and XXZ spin chains 
have been derived by the fusion method in several references 
\cite{FusionXXX,Babujian,Babujian-Tsvelick,Fateev-Zamolodchikov,SAA,KR,V-DWA}. 
In the region: $0 \le \zeta < \pi/2s$,  
the spin-$s$ ground-state should be given 
by a set of string solutions \cite{Takhtajan,Sogo}. 
Furthermore, the critical behavior should be 
given by the SU(2) WZWN model of level $k=2s$ with central charge 
$c=3s/(s+1)$ 
\cite{Johannesson,Johannesson2,Alcaraz-Martins:XXX,Affleck,Dorfel,Avdeev,Alcaraz-Martins:XXZ,Fowler,Frahm,deVega-Woynarovich,KR,KB,KBP,JSuzuki}. 
For the integrable higher-spin XXZ spin chain 
correlation functions have been discussed in the massive regime 
by the method of $q$-vertex operators 
\cite{Idzumi0,Idzumi2,Bougourzi,Konno}.

In the present paper we discuss several important points 
in the algebraic Bethe-ansatz derivation of 
the correlation functions 
for the integrable spin-$s$ XXZ spin chain 
where $s$ is an arbitrary integer or a half-integer \cite{DM2}. 
In particular, we briefly discuss a rigorous derivation of 
the finite-sum expression of correlation functions 
for the spin-$s$ XXZ spin chain.  

The content of the paper consists of the following. 
In section 2 we formulate the $R$-matrices in the homogeneous and 
principal gradings, respectively. They are related to each other 
by a similarity transformation.   In section 3 we introduce 
the Hermitian elementary matrices 
and construct conjugate basis vectors for the spin-$s$ Hilbert space 
in the massless regime.  
In section 4 we construct fusion monodromy matrices. 
In section 5, we first present formulas \cite{DM2} for expressing 
the Hermitian elementary matrices in terms of global operators. 
Then, we review the multiple-integral representations 
of the spin-$s$ XXZ correlation function 
for an arbitrary product of elementary matrices. \cite{DM2}     
In section 6 we briefly sketch the derivation of 
the finite-sum expression of correlation functions 
for the spin-$s$ XXZ spin chain, 
which leads to the multiple-integral 
representation in the thermodynamic limit.  
Here the spin-1/2 case corresponds to 
eq. (5.6) of Ref. \refcite{KMT2000}.

%
%
\setcounter{section}{1} 
\setcounter{equation}{0} 
 \renewcommand{\theequation}{2.\arabic{equation}}
\section{Symmetric and asymmetric $R$-matrices }

\subsection{$R$-matrix and the monodromy matrix of type $(1,1^{\otimes L})$}

Let us now define the $R$-matrix of the XXZ spin chain.  
\cite{Korepin,MS2000,KMT1999,KMT2000}  
For two-dimensional vector spaces $V_1$ and $V_2$,  
we define ${R}^{\pm}(\lambda_1-\lambda_2)$ 
acting on $V_1 \otimes V_2$ by 
\begin{equation} 
{R}^{\pm}(\lambda_1-\lambda_2) = \sum_{a,b,c,d=0,1} 
R^{\pm}(u)^{a \, b}_{c \, d} \, 
\, e^{a, \, c} \otimes e^{b, \, d} = 
\left( 
\begin{array}{cccc}
1 & 0 & 0 & 0 \\ 
0 & b(u) & c^{\mp}(u) & 0 \\
0 & c^{\pm}(u) & b(u) & 0 \\
0 & 0 & 0 & 1 \\
\end{array} 
\right) \, \label{eq:R+},  
\end{equation}
where $u=\lambda_1-\lambda_2$, 
$b(u) = \sinh u/\sinh(u + \eta)$ and 
$c^{\pm}(u) = \exp( \pm u) \sinh \eta/\sinh(u + \eta)$. 
We denote by $e^{a, \, b}$ a unit matrix that has only one nonzero element 
equal to 1 at entry $(a, b)$ where $a, b= 0, 1$.

The asymmetric $R$-matrix (\ref{eq:R+}), $R^{+}(u)$,  
is compatible with the homogeneous grading 
of $U_q(\widehat{sl}_2)$ \cite{Jimbo-Miwa,DM1}. 
We denote by $R^{(p)}(u)$ or simply by $R(u)$ the symmetric $R$-matrix 
where $c^{\pm}(u)$ of (\ref{eq:R+}) are replaced by 
$c(u)= \sinh \eta/\sinh(u+\eta)$ \cite{DM1}. 
It is compatible with the affine quantum group $U_q(\widehat{sl}_2)$ 
of the principal grading \cite{Jimbo-Miwa,DM1}.
Hereafter, we denote them concisely by $R^{(w)}(u)$ 
with $w= \pm$ and $p$,  where $w=+$ and $w=p$ in superscript 
show the homogeneous and the principal grading, respectively

Let $s$ be an integer or a half-integer.  We shall mainly 
consider the tensor product 
$V_1^{(2s)} \otimes \cdots \otimes V_{N_s}^{(2s)}$ 
of $(2s+1)$-dimensional vector spaces $V^{(2s)}_j$ with 
parameters $\xi_j$, where $L= 2s N_s$. 
Here $N_s$ denotes the lattice size of the spin-$s$ chain. 
In general, we may consider the tensor product 
$V_0^{(2s_0)} \otimes V_1^{(2s_1)} \otimes \cdots 
\otimes V_{r}^{(2s_{r})}$ with $2s_1 + \cdots + 2s_r = L$,  
where $V_{j}^{(2s_j)}$ have parameters 
$\lambda_j$ or $\xi_j$  for $j=1, 2, \ldots, r$.    
For a given set of matrix elements 
$A^{a, \, \alpha}_{b, \, \beta}$  for $a,b=0, 1, \ldots, 2s_j$ and  
$\alpha, \beta=0, 1, \ldots, 2s_k$,  
we define operator $A_{j, k}$ by 
\begin{eqnarray} 
A_{j, k} & = & \sum_{a,b=1}^{\ell} \sum_{\alpha, \beta} 
A^{a, \, \alpha}_{b, \, \beta} I_0^{(2s_0)} \otimes
I_1^{(2s_1)} \otimes \cdots \otimes I_{j-1}^{(2s_{j-1})}  
\otimes E_j^{a, b \, (2s_j)} \otimes \nonumber\\ 
&   
\otimes & I_{j+1}^{(2s_{j+1})}  \otimes \cdots \otimes I_{k-1}^{(2s_{k-1})}  
\otimes E_k^{\alpha, \beta \, (2s_k)}  \otimes I_{k+1}^{(2s_{k+1})} 
 \otimes \cdots \otimes I_{r}^{(2s_r)} .  \nonumber \\
\label{defAjk}
\end{eqnarray}
Here $E_j^{a, b \, (2s_j)}$ denote the elementary matrices 
in the spin-$s_j$ representation, each of which has nonzero 
matrix element only at entry $(a,b)$. 

When $s_0=\ell/2$ and $s_1=\cdots s_r= s$, 
we denote the type by $(\ell, (2s)^{\otimes N_s})$. 
In particular, for $s=1/2$, we denote it by $(\ell, 1^{\otimes L})$.

%
%
%
\subsection{Gauge transformations}

Let us introduce operators ${\Phi}_j$ 
with arbitrary parameters $\phi_j$ for $j=0, 1, \ldots, L$ as follows:  
\begin{equation} 
{\Phi}_j = \left( 
\begin{array}{cc}  
1 & 0 \\
0 & e^{\phi_j} 
\end{array}
\right)_{[j]} =  
I^{\otimes (j)} \otimes \left( 
\begin{array}{cc}  
1 & 0 \\
0 & e^{\phi_j} 
\end{array}
\right) \otimes I^{\otimes (L-j)}  . 
\label{eq:Phi}
\end{equation}
In terms of ${\chi}_{jk}= {\Phi}_j {\Phi}_k$,  
we define a similarity transformation on the $R$-matrix by  
\begin{equation} 
R_{jk}^{\chi} = {\chi}_{jk} R_{jk} {\chi}_{jk}^{-1} \, . 
\end{equation}
Explicitly, the following two matrix elements are transformed. 
\begin{equation} 
\left( R_{jk}^{\chi} \right)^{21}_{12} = 
c(\lambda_j, \lambda_k) e^{\phi_j-\phi_k} \, , \quad 
\left( R_{jk}^{\chi} \right)^{12}_{21} = 
c(\lambda_j, \lambda_k) e^{-\phi_j+\phi_k} \, . 
\end{equation}
Putting $\phi_j= \lambda_j$ for $j=0, 1, \ldots, L$ 
in eq. (\ref{eq:Phi}) we have 
\begin{equation} 
R_{jk}^{\pm}(\lambda_j, \lambda_k) = \left( \chi_{jk} \right)^{\pm 1}  
\, R_{jk}(\lambda_j, \lambda_k) \, 
\left( \chi_{jk} \right)^{\mp 1} 
\quad (j, k = 0, 1, \ldots, L) .  
\end{equation}
Thus, the asymmetric $R$-matrices $R_{12}^{\pm}(\lambda_1, \lambda_2)$  
are derived from the symmetric one through the gauge transformation 
$\chi_{jk}$.

\subsection{Monodromy matrices}

Applying definition (\ref{defAjk}) 
for matrix elements $R(u)^{ab}_{cd}$ of a given $R$-matrix, 
$R^{(w)}(u)$ for $w= \pm$ and $p$,  
we define  $R$-matrices 
$R_{j k}^{(w)}(\lambda_j, \lambda_k)=R_{j k}^{(w)}(\lambda_j-\lambda_k)$ 
 for integers $j$ and $k$ with $0\le j < k \le L$. 
For integers $j, k$ and $\ell$ with $0 \le j < k < \ell \le L$, 
the $R$-matrices satisfy the Yang-Baxter equations 
\begin{eqnarray}  
& &  R_{j k}^{(w)}(\lambda_j-\lambda_k) 
R_{j \ell}^{(w)}(\lambda_j-\lambda_{\ell})
R_{k \ell}^{(w)}(\lambda_k-\lambda_{\ell}) 
\nonumber\\ 
& = & 
 R_{k \ell}^{(w)}(\lambda_k-\lambda_{\ell}) 
R_{j \ell}^{(w)}(\lambda_j-\lambda_{\ell})
R_{j k}^{(w)}(\lambda_j-\lambda_k) \, . \label{eq:YBE}
\end{eqnarray}

Let us introduce notation  for expressing products of $R$-matrices. 
\begin{eqnarray} 
R_{1, 23 \cdots n}^{(w)} & = & R_{1 n}^{(w)}  
\cdots R_{13}^{(w)} R_{12}^{(w)}  \, , \nonumber\\
R_{12 \cdots n-1, n}^{(w)}  & = & 
R_{1 n}^{(w)}  R_{2 n}^{(w)}  \cdots R_{n-1 \, n}^{(w)}  \, . 
\label{eq:useful-notation}
\end{eqnarray}
Here $R_{a b}^{(w)} $ denote the $R$-matrix 
$R_{a b}^{(w)} =R_{a b}^{(w)} (\lambda_a-\lambda_b)$ 
for $a, b =1, 2, \ldots, n$.   

We now define the monodromy matrix of type $(1, 1^{\otimes L} \, w)$, i.e. 
of type $(1,1^{\otimes L})$ with grading $w$.  
Expressing the symbol $(1,1^{\otimes L})$ 
briefly as $(1, \, 1)$ in superscript we define it by    
\begin{eqnarray} 
& & T_{0, \, 1 2 \cdots L}^{(1, \, 1 \, w)}(\lambda_0; \{w_j \}_L)  
= R_{0L}^{+}(\lambda_0-w_L) 
 \cdots R_{02}^{+}(\lambda_0-w_2) R_{01}^{+}(\lambda_0-w_1) 
\nonumber \\ 
&  & \quad = R_{0 L}^{(w)} R_{0 L-1}^{(w)} \cdots R_{0 1}^{(w)} 
\,  = R_{0, \, 1 2 \cdots L}^{(\, w)}(\lambda_0; \{w_j \}_L) 
 \, . 
\end{eqnarray}
Here we have put $\lambda_j=w_j$ for $j=1, 2, \ldots, L$. 
They are arbitrary. 
We call them {\it inhomogeneous parameters}.    
We express the operator-valued matrix elements of 
the monodromy matrix as follows.    
\begin{equation}
T^{(1, 1 \, +)}_{0, 1 2 \cdots L}(\lambda; \{ w_j \}_L ) = 
\left( 
\begin{array}{cc} 
A^{(1 +)}_{1 2 \cdots L}(\lambda; \{ w_j \}_L) & 
B^{(1 +)}_{1 2 \cdots L}(\lambda; \{ w_j \}_L) \\ 
C^{(1 +)}_{1 2 \cdots L}(\lambda; \{ w_j \}_L) & 
D^{(1 +)}_{1 2 \cdots L}(\lambda; \{ w_j \}_L)  
\end{array} 
\right) \, . 
\end{equation}
 We also denote the operator-valued matrix elements by   
$[T^{(1, 1 +)}_{0, 1 2 \cdots L}(\lambda; \{ w_j \}_L )]_{a,b}$ 
for $a,b=0,1$. Here $\{ w_j \}_L$ denotes the inhomogeneous parameters 
 $w_1, w_2, \ldots, w_L$. Hereafter we denote 
by $\{ \mu_j \}_N$ the set of $N$ numbers or parameters 
$\mu_1, \ldots, \mu_N$. 

The monodromy matrix of principal 
grading, $T^{(1, 1 \, p)}_{0, 1 2 \cdots L}(\lambda; \{ w_j \}_L )$,    
is related to that of homogeneous grading 
via similarity transformation 
$\chi_{0 1 \cdots L}= \Phi_0 \Phi_1 \cdots \Phi_L$  as follows \cite{DM1}.  
 \begin{eqnarray}
& & T^{(1, 1 \, +)}_{0, 1 2 \cdots L}(\lambda; \{ w_j \}_L )  =  
\chi_{0 1 2 \cdots L} 
T^{(1, 1 \, p)}_{0, 1 2 \cdots L}(\lambda; \{ w_j \}_L )
\chi_{0 1 2 \cdots L}^{-1}  
\nonumber\\ 
&  & = 
\left( 
\begin{array}{cc} 
\chi_{1 2 \cdots L} 
A^{(1 \,p)}_{1 2 \cdots L}(\lambda; \{ w_j \}_L) 
\chi_{1 2 \cdots L}^{-1} & 
e^{- \lambda_0} \chi_{1 2 \cdots L} 
B^{(1 \, p)}_{1 2 \cdots L}(\lambda; \{ w_j \}_L) 
\chi_{1 2 \cdots L}^{-1}  \\ 
e^{\lambda_0}
\chi_{1 2 \cdots L} 
C^{(1 \, p)}_{1 2 \cdots L}(\lambda; \{ w_j \}_L) 
\chi_{1 2 \cdots L}^{-1}  & 
\chi_{1 2 \cdots L} D^{(1 \, p)}_{1 2 \cdots L}(\lambda; \{ w_j \}_L) 
\chi_{1 2 \cdots L}^{-1}   
\end{array} 
\right) \, . \nonumber 
\end{eqnarray}
In Ref. \cite{DM1} operator 
$A^{(1 \, +)}(\lambda)$ has been written as  $A^{+}(\lambda)$.

\subsection{Operator ${\check R}$: Another form of the $R$-matrix}

Let  $V_1$ and $V_2$ be $(2s+1)$-dimensional vector spaces.     
We define permutation operator $\Pi_{1, \, 2}$ by   
\begin{equation} 
\Pi_{1, \, 2} \, v_1 \otimes v_2 = 
v_2 \otimes v_1 \, , \quad v_1 \in V_1 \, , \, v_2 \in V_2 \, .  
\label{eq:Pi-2s-2s}
\end{equation}
In the spin-1/2 case,  
we define operator ${\check R}_{j \, j+1}^{(w)}(u)$ by 
\begin{equation} 
{\check  R}_{j ,  \, j+1}^{(w)}(u)
= \Pi_{j, \, j+1} \,  R_{j, \, j+1}^{(w)}(u) \, . 
\label{eq:Rcheck-spin-1/2}
\end{equation}

 \setcounter{equation}{0} 
 \renewcommand{\theequation}{3.\arabic{equation}} 
\section{The quantum group invariance}
\subsection{Quantum group $U_q(sl_2)$}

The quantum algebra $U_q(sl_2)$ 
is an associative algebra over ${\bf C}$ generated by  
$X^{\pm}, K^{\pm}$  with the following relations: \cite{Jimbo-QG,Jimbo-Hecke,Drinfeld}
\begin{eqnarray} 
K K^{-1} & = & K K^{-1} = 1 \, , \quad 
K X^{\pm} K^{-1}  =  q^{\pm 2} X^{\pm} \, ,  \quad 
\, , \nonumber \\
{[} X^{+}, X^{-} {]} & = &  
{\frac   {K - K^{-1}}  {q- q^{-1}} } \, . 
\end{eqnarray}
The algebra $U_q(sl_2)$ is also a Hopf algebra over ${\bf C}$ 
with comultiplication 
\begin{eqnarray} 
\Delta (X^{+}) & = & X^{+} \otimes 1 + K \otimes X^{+}  \, , 
 \quad 
\Delta (X^{-})  =  X^{-} \otimes K^{-1} + 1 \otimes X^{-} \, ,  \nonumber \\
\Delta(K) & = & K \otimes K  \, , 
\end{eqnarray} 
and antipode:  
$S(K)=K^{-1} \, , S(X^{+})= - K^{-1} X^{+} \, , S(X^{-}) = -  X^{-} K$, and   
coproduct: $\epsilon(X^{\pm})=0$ and $\epsilon(K)=1$.

It is easy to see that the asymmetric $R$-matrix gives an intertwiner 
of the spin-1/2 representation of $U_q(sl_2)$: 
\begin{equation} 
R_{12}^{+}(u) \Delta (x) = 
\tau \circ \Delta (x)  R^{+}_{12}(u) 
\quad {\rm for \, \,} \quad x= X^{\pm}, K \, .  
\label{eq:RD}
\end{equation}
Here we remark that spectral parameter $u$ is arbitrary and 
independent of $X^{\pm}$ or $K$. 
%

%
%
\subsection{Temperley-Lieb algebra} 

Operators ${\check R}_{j, j+1}^{\pm}(u)$ are decomposed  
in terms of the generators of the Temperley-Lieb algebra as follows 
\cite{Baxter-book}. 
\begin{equation} 
{\check R}_{j,j+1}^{\pm}(u) = I - b(u) U_{j}^{\pm} \, . 
\label{eq:TLdecomp+}
\end{equation} 
$U_j^{+}$s  ($U_j^{-}$s) 
satisfy the defining relations of the Temperley-Lieb algebra: 
\cite{Baxter-book}
\begin{eqnarray} 
U^{\pm}_{j} U^{\pm}_{j + 1} U^{\pm}_j & = & U^{\pm}_j, \nonumber \\  
U^{\pm}_{j+1} U^{\pm}_{j } U^{\pm}_{j+1} & = & U^{\pm}_j, 
\quad {\rm for} \, j=0, 1, \ldots, L-2, 
\nonumber \\ 
\left( U^{\pm}_j \right)^2 
& = &  (q+q^{-1}) \, U^{\pm}_j \quad {\rm for} 
\, j=0, 1, \ldots, L-1, 
\nonumber \\  
U^{\pm}_j U^{\pm}_k & = & U^{\pm}_k U^{\pm}_j 
\quad {\rm for} \, \, |j-k| > 1\, .  
\end{eqnarray}

We remark that the asymmetric $R$-matrices 
${\check R}^{\pm}_{j, j+1}(u)$ 
derived from the symmetric $R$-matrix through the gauge transformation 
are related to  the Jones polynomial. \cite{AW}

%
%
\subsection{Basis vectors of spin-$\ell/2$ representation of $U_q(sl_2)$}

Let us introduce 
  the $q$-integer for an integer $n$ by 
$[n]_q= (q^n-q^{-n})/(q-q^{-1})$.  
We define the $q$-factorial  $[n]_q!$ for integers $n$ by 
\begin{equation} 
[n]_q ! = [n]_q \, [n-1]_q \, \cdots \, [1]_q \, .  
\end{equation}
For integers $m$ and $n$ satisfying $m \ge n \ge 0$ 
we define the $q$-binomial coefficients as follows
\begin{equation} 
\left[ 
\begin{array}{c} 
m \\ 
n 
\end{array}  
 \right]_q 
= {\frac {[m]_q !} {[m-n]_q ! \, [n]_q !}}  \, . 
\end{equation} 

We now define the basis vectors of the $(\ell+1)$-dimensional 
irreducible representation of $U_q(sl_2)$,  
$|| \ell, n  \rangle$ for $n=0, 1, \ldots, \ell$ as follows. 
We define $||\ell, 0 \rangle$ by 
\begin{equation} 
||\ell , 0 \rangle = |0 \rangle_1 \otimes |0 \rangle_2 \otimes 
\cdots \otimes |0 \rangle_\ell \, . 
\end{equation}   
Here  $|\alpha \rangle_j$ for $\alpha=0, 1$ 
denote the basis vectors of the spin-1/2 representation defined 
on the $j$th position in the tensor product.  We define 
$|| \ell, n \rangle$ for $n \ge 1$ and evaluate them as follows \cite{DM1} .   
\begin{eqnarray} 
|| \ell, n \rangle  & =  &  
\left( \Delta^{(\ell-1)} (X^{-}) \right)^n ||\ell, 0 \rangle \,  
{\frac 1 {[n]_q!}} \nonumber \\ 
%
& = &
\sum_{1 \le i_1 < \cdots < i_n \le \ell} 
\sigma_{i_1}^{-} \cdots \sigma_{i_n}^{-} | 0 \rangle \, 
q^{i_1+ i_2 + \cdots + i_n  - n \ell + n(n-1)/2} \, . 
\label{eq:|ell,n>} 
\end{eqnarray}
We define the conjugate vectors explicitly by the following: 
\begin{equation} 
\langle \ell, n || =  
\left[ 
\begin{array}{c} 
\ell \\ 
n 
\end{array}  
 \right]_q^{-1} \, q^{n(\ell-n)} \, 
\sum_{1 \le i_1 < \cdots < i_n \le \ell} 
\langle 0 | \sigma_{i_1}^{+} \cdots \sigma_{i_n}^{+} \, 
q^{i_1 + \cdots + i_n - n \ell + n(n-1)/2}   \, . 
\label{eq:<ell,n|}
\end{equation}
It is easy to show the normalization conditions \cite{DM1}: 
$\langle \ell, n || \, || \ell, n \rangle = 1$.  
In the massive regime where $q = \exp \eta$ with real $\eta$, 
conjugate vectors $\langle \ell, n || $ are Hermitian conjugate to vectors 
$|| \ell, n \rangle$.

%
%
\subsection{Conjugate vectors } 

In order to construct Hermitian elementary matrices 
in the massless regime where $|q|=1$, 
we now introduce another set of dual basis vectors.   
For a given nonzero integer $\ell$ we define  
$\widetilde{\langle \ell, n ||}$ for $n=0, 1, \ldots, n$, by 
\begin{equation} 
\widetilde{\langle \ell, n ||} =  
\left( 
\begin{array}{c} 
\ell \\ 
n 
\end{array}  
 \right)^{-1} \,  
\sum_{1 \le i_1 < \cdots < i_n \le \ell} 
\langle 0 | \sigma_{i_1}^{+} \cdots \sigma_{i_n}^{+} \, 
q^{-(i_1 + \cdots + i_n) + n \ell - n(n-1)/2}   \, . 
\label{eq:tilde<ell,n|}
\end{equation}
They are conjugate to $|| \ell, n \rangle$:   
$\widetilde{\langle \ell, m ||} \, || \ell, n \rangle = 
\delta_{m, n} $ . 
Here we have denoted the binomial coefficients 
for integers $\ell$ and $n$ with $0 \le n \le \ell$ as follows. 
\begin{equation} 
\left( 
\begin{array}{cc} 
\ell \\ 
n 
\end{array} 
 \right)
= {\frac {\ell !} {(\ell-n)! n!}} \, .  
\end{equation}

We now introduce vectors $\widetilde{|| \ell, n \rangle}$ 
which are Hermitian conjugate to $\langle \ell, n ||$ when 
$|q|=1$ 
for positive integers $\ell$ with $n=0, 1, \ldots, \ell$. 
Setting the norm of $\widetilde{|| \ell, n \rangle}$ 
such that 
$\langle \ell, n || \,  \widetilde{|| \ell, n \rangle}=1$, 
vectors $\widetilde{|| \ell, n \rangle}$ are given by   
\begin{equation} 
\sum_{1 \le i_1 < \cdots < i_n \le \ell} \sigma_{i_1}^{-} 
\cdots \sigma_{i_n}^{-} | 0 \rangle 
q^{-(i_1 + \cdots + i_n) + n \ell - n(n-1)/2} 
\left[ 
\begin{array}{cc} 
\ell \\ 
n 
\end{array} 
 \right]_q \, 
q^{-n(\ell-n)} 
\left( 
\begin{array}{cc} 
\ell \\ 
n 
\end{array} 
 \right)^{-1} \, . 
\end{equation}

We have the following normalization condition: 
\begin{equation}
\widetilde{\langle \ell, n ||} \, \widetilde{|| \ell, n \rangle} 
= \left[ 
\begin{array}{cc} 
\ell \\ 
n 
\end{array} 
 \right]_q^2 \,  
\left( 
\begin{array}{cc} 
\ell \\ 
n 
\end{array} 
 \right)^{-2} \, .  
\end{equation}

\subsection{Hermitian elementary matrices}

In the massless regime we define elementary matrices 
${\widetilde E}^{m, \, n \, (2s \, + )}$ for $m, n=0, 1, \ldots, 2s$ 
by 
\begin{equation} 
{\widetilde E}^{m, \, n \, (2s \, + )}
= \widetilde{||2s, m \rangle} \, \langle 2s, n || \,  . 
\end{equation}

In the massless regime where $|q|=1$,  
matrix $|| \ell, n \rangle  \widetilde{\langle \ell, n ||}$ is 
Hermitian: $(|| \ell, n \rangle  \widetilde{\langle \ell, n ||})^{\dagger} 
= || \ell, n \rangle  \widetilde{\langle \ell, n ||}$. 
However, in order to define projection operators $\tilde{P}$ such that 
$P \tilde{P} = P$, we have formulated  vectors 
 $\widetilde{|| \ell, n \rangle}$.

%
%
\subsection{Projection operators} 

We define projection operators acting on the 1st to 
the $\ell$th tensor-product spaces by     
\begin{equation} 
P^{(\ell)}_{1 2 \cdots \ell} 
= \sum_{n=0}^{\ell}  || \ell, n \rangle \, \langle \ell, n || \, . 
\label{eq:Psum}
\end{equation} 
Let us now introduce another set of projection operators 
$\widetilde{P}_{1 \cdots \ell}^{(\ell)}$ 
as follows. 
\begin{equation}
\widetilde{ P}_{1 \cdots \ell}^{(\ell)} = \sum_{n=0}^{\ell} 
\widetilde{ || \ell , \,  n \rangle} \langle \ell , \, n ||  \, . 
\label{eq:P'sum}
\end{equation}
Projector $\widetilde{ P}_{1 \cdots \ell}^{(\ell)}$ is idempotent: 
$(\widetilde{ P}_{1 \cdots \ell}^{(\ell)})^2= 
\widetilde{ P}_{1 \cdots \ell}^{(\ell)}$.   
In the massless regime where $|q|=1$, 
it is Hermitian:  
$\left( \widetilde{ P}_{1 \cdots \ell}^{(\ell)} \right)^{\dagger}= 
\widetilde{ P}_{1 \cdots \ell}^{(\ell)}$.  
From (\ref{eq:Psum}) and (\ref{eq:P'sum}), 
we show the following properties: 
\begin{eqnarray} 
P_{1 2 \cdots \ell}^{(\ell)} 
\widetilde{ P}_{1 \cdots \ell}^{(\ell)}
& = & P_{1 2 \cdots \ell}^{(\ell)} \, ,  
\label{eq:PP'=P} \\
\widetilde{ P}_{1 \cdots \ell}^{(\ell)}
P_{1 2 \cdots \ell}^{(\ell)} 
& = & \widetilde{ P}_{1 \cdots \ell}^{(\ell)} \, . 
\label{eq:P'P=P'}
\end{eqnarray}
In the tensor product of quantum spaces, 
$V^{(2s)}_1 \otimes \cdots \otimes V_{N_s}^{(2s)}$,   
 we define $\widetilde{P}_{12 \cdots L}^{(2s)}$ by 
\begin{equation}  
\widetilde{P}_{12 \cdots L}^{(2s)} 
= \prod_{i=1}^{N_s} \widetilde{P}^{(2s)}_{2s(i-1)+1}  \, . 
\end{equation}
Here we recall $L = 2s N_s $.

The projection operators are also constructed by the fusion method. 
 For $\ell > 2$ we can construct projection operators  
 inductively with respect to $\ell$ as follows 
\cite{Jimbo-Hecke,V-DWA}. 
\begin{equation} 
P_{1 2 \cdots \ell}^{(\ell)} = 
P_{1 2 \cdots \ell-1}^{(\ell-1)} {\check R}^{+}_{\ell-1, \, \ell}
((\ell-1)\eta) P_{12\cdots \ell-1}^{(\ell-1)} \, . 
\label{eq:def-projector}
\end{equation} 
The projection operator $P_{12\cdots \ell}^{(\ell)}$ 
gives a $q$-analogue of the full symmetrizer 
of the Young operators for the Hecke algebra \cite{Jimbo-Hecke}.

%
%
\setcounter{equation}{0} 
 \renewcommand{\theequation}{4.\arabic{equation}}
%
%
\section{Fusion construction}

\subsection{Higher-spin monodromy matrix of type 
$(\ell, \, (2s)^{\otimes N_s})$ }

We now set the inhomogeneous parameters $w_j$ for $j=1, 2, \ldots, L$,  
as $N_s$ sets of complete $2s$-strings \cite{DM1}. 
We define $w_{(b-1)\ell+ \beta}^{(2s)}$ for $\beta = 1, \ldots, 2s$, 
as follows.  
\begin{equation} 
w_{2s(b-1)+ \beta}^{(2s)} = \xi_b - (\beta-1) \eta \, , \quad 
 \mbox{for} \quad b = 1, 2, \ldots, N_s . 
\label{eq:2s-strings}
\end{equation}
We shall define the monodromy matrix of type $(1, (2s)^{\otimes N_s})$ 
associated with homogeneous grading.  
We first define the massless monodromy matrix by  
\begin{eqnarray} 
\widetilde{T}^{(1, \, 2s \, +)}_{0, \, 1 2 \cdots N_s}
(\lambda_0; \{ \xi_b \}_{N_s} ) 
& = & \widetilde{P}_{12 \cdots L}^{(2s)} 
R_{0, \, 1 \ldots L}^{(1, \, 1 \, +)} 
(\lambda_0; \{ w_{j}^{(2s)} \}_L) 
\widetilde{P}_{12 \cdots L}^{(2s)}  \nonumber \\  
& = & 
\left( 
\begin{array}{cc} 
\widetilde{A}^{(2s +)}(\lambda; \{ \xi_b \}_{N_s}) & 
\widetilde{B}^{(2s +)}(\lambda; \{ \xi_b \}_{N_s}) \\ 
\widetilde{C}^{(2s +)}(\lambda; \{ \xi_b \}_{N_s}) & 
\widetilde{D}^{(2s +)}(\lambda; \{ \xi_b \}_{N_s})  
\end{array} 
\right) \, . 
\end{eqnarray}

Let us introduce a set of $2s$-strings with small deviations from 
the set of complete $2s$-strings.   
\begin{eqnarray} 
w_{2s(b-1)+ \beta}^{(2s; \, \epsilon)}& = & \xi_b - (\beta-1) \eta  
+ \epsilon r_b^{(\beta)} \, , \quad 
 \mbox{for} \quad b=1, 2, \cdots, N_s, \, \nonumber \\ 
& & \qquad \qquad \qquad \mbox{and} \, \, \beta=1, 2, \ldots, 2s. 
\label{eq:2s-strings-epsilon}
\end{eqnarray}
Here $\epsilon$ is very small and 
$r_{b}^{(\beta)}$ are generic parameters.  
We express the elements of the monodromy matrix $T^{(1,1)}$ with 
inhomogeneous parameters given by $w_j^{(2s; \, \epsilon)}$ 
for $j=1, 2, \ldots, L$ as follows. 
\begin{equation}
T^{(1, \, 1 \, +)}_{0, \, 1 2 \cdots L}
(\lambda; \{ w_j^{(2s; \epsilon)} \}_{L}) = 
\left( 
\begin{array}{cc} 
A^{(2s +; \, \epsilon)}_{1 2 \cdots L}(\lambda) & 
B^{(2s +; \, \epsilon)}_{1 2 \cdots L}(\lambda) \\ 
C^{(2s +; \, \epsilon)}_{1 2 \cdots L}(\lambda) & 
D^{(2s +; \, \epsilon)}_{1 2 \cdots L}(\lambda)  
\end{array} 
\right) \, . 
\end{equation}
Here 
$A^{(2s + ;  \, \epsilon)}_{1 2 \cdots L}(\lambda)$ denotes 
$A^{(1+)}_{1 2 \cdots L}(\lambda; \{ w_j^{(2s;  \, \epsilon)} \}_L)$. 
\begin{equation} 
\widetilde{ A}^{(2s +)}_{1 2 \cdots N_s}(\lambda; \{\xi_p \}_{N_s}) = 
\lim_{\epsilon \rightarrow 0} 
\widetilde{P}_{1 2 \cdots L}^{(2s)}  
A^{(2s +; \, \epsilon)}_{1 2 \cdots L}
(\lambda; \{w_j^{(2s; \, \epsilon)} \}_{L}) 
\widetilde{P}_{1 2 \cdots L}^{(2s)}  \, . 
\end{equation}

We define the massless monodromy matrix of type 
$(\ell, \, (2s)^{\otimes N_s})$ by 
\begin{eqnarray} 
\widetilde{T}^{(\ell, \, 2s \, +)}_{0, \, 1 2 \cdots N_s} 
& = & \widetilde{P}^{(\ell)}_{a_1 a_2 \cdots a_{\ell}} \,   
\widetilde{T}_{a_1, \, 1 2 \cdots N_s}^{(1, \, 2s \, +)}(\lambda_{a_1}) 
\widetilde{T}_{a_2, \, 1 2 \cdots N_s}^{(1, \, 2s \, +)}(\lambda_{a_1}-\eta) 
\cdots \nonumber \\ 
& & \times \cdots 
\widetilde{T}_{a_{\ell}, \, 1 2 \cdots N_s}^{(1, \,  2s \, +)}
(\lambda_{a_1}-(\ell-1)\eta) \,  
\widetilde{P^{(\ell)}}_{a_1 a_2 \cdots a_{\ell}} \, .  
\end{eqnarray}

%
%
\subsection{Integrable spin-$s$ Hamiltonians}

We define the massless transfer matrix \cite{DM2} of type 
$(\ell, (2s)^{\otimes N_s})$ by 
\begin{eqnarray} 
& & \widetilde{t}^{(\ell, \, 2s \, +)}_{1 2 \cdots N_s}(\lambda) 
= {\rm  tr}_{V^{(\ell)}}
\left( \widetilde{T}^{(\ell, \, 2s \, +)}_{0, \, 1 2 \cdots N_s}
(\lambda) \right) 
 = \sum_{n=0}^{\ell} {}_a \langle \ell,  n || \, 
 \widetilde{T}^{(1, \, 2s \, +)}_{a_1, \, 1 2 \cdots N_s}(\lambda) 
\, \times \nonumber \\ 
&  & \times \, 
\widetilde{T}^{(1, \, 2s \, +)}_{a_2, \, 1 2 \cdots N_s}(\lambda-\eta)  
\cdots \widetilde{T}^{(1, \, 2s \, +)}_{a_{\ell}, \, 1 2 \cdots N_s}
(\lambda-(\ell-1) \eta) \, 
\widetilde{ || \ell, n \rangle}_a  \, .   
\end{eqnarray}
It follows from the Yang-Baxter equations that 
the higher-spin transfer matrices commute in the tensor product space 
$V_1^{(2s)} \otimes \cdots \otimes V_{N_s}^{(2s)}$, 
which is derived by applying 
projection operator $P^{(2s)}_{1 2 \cdots L}$ 
to $V^{(1)}_1 \otimes \cdots \otimes V_L^{(1)}$.

The massless spin-$s$ $R$-matrix 
$\widetilde{R}_{1 \, 2}^{(2s, \, 2s \, +)}(u)$
becomes the permutation operator at $u=0$: 
$\widetilde{R}_{1 \, 2}^{(2s, \, 2s \, +)}(0) = \Pi_{1, \, 2}$. 
\cite{Jimbo-QG,DMo} 
Therefore, 
putting inhomogeneous parameters $\xi_p=0$ for $p=1, 2, \ldots, N_s$, 
we show that the transfer matrix 
$\widetilde{t}^{(2s, \, 2s \, +)}_{1 2 \cdots N_s}(\lambda)$ 
becomes the shift operator at $\lambda=0$. 
We  derive the massless spin-$s$ XXZ Hamiltonian 
by the logarithmic derivative of the massless spin-$s$ transfer matrix.   
\begin{eqnarray} 
{\cal H}^{(2s)}_{\rm XXZ} = \left. {\frac d {d \lambda}} 
\log \widetilde{t}^{(2s,  \, 2s \, +)}_{1 2 \cdots N_s}(\lambda)
\right|_{\lambda=0 , \, \xi_j=0} 
= 
\sum_{i=1}^{N_s} 
\left. \frac d {du} 
\widetilde{\check R}_{i, i+1}^{(2s, 2s)}(u) \right|_{u=0}  \, . 
\label{eq:deriv-XXZ-Hamiltonian}
\end{eqnarray}

%
%
%
%
 \setcounter{equation}{0} 
 \renewcommand{\theequation}{5.\arabic{equation}}
%
%
\section{Spin-$\ell/2$ massless XXZ correlation functions}

\subsection{Spin-$s$ local operators  
in terms of global operators}
%

In the massless regime, we can express the Hermitian elementary 
matrices in terms of global operators as follows.\cite{DM2} 
For $m \ge n$ we have 
\begin{eqnarray}  
& & \widetilde{E}_{i}^{m, \, n \, (\ell \, + )} 
 = \left(
\begin{array}{c} 
\ell \\
n 
\end{array} 
 \right) \, 
\left[
\begin{array}{c} 
\ell \\
m 
\end{array} 
 \right]_q \, 
\left[
\begin{array}{c} 
\ell \\
n 
\end{array} 
 \right]_q^{-1} \, 
\widetilde{P}^{(\ell)}_{1 \cdots L} 
\,  
\prod_{\alpha=1}^{(i-1) \ell} (A^{(1+)}+D^{(1+)})(w_{\alpha}) 
\nonumber \\ 
& & \times 
\prod_{k=1}^{n} D^{(1+)}(w_{(i-1)\ell+k} ) 
 \prod_{k=n+1}^{m} B^{(1+)}(w_{(i-1)\ell + k})  
\, \prod_{k=m+1}^{\ell} A^{(1+)}(w_{(i-1) \ell+k})  
\nonumber \\ 
& &  \times
\prod_{\alpha=i \ell +1}^{\ell N_s} 
(A^{(1+)}+D^{(1+)})(w_{\alpha}) \, \,   
\widetilde{P}^{(\ell)}_{1 \cdots L}
\, .  \label{eq:Em>n}
\end{eqnarray}
For $m \le n$ we have 
\begin{eqnarray}  
& & \widetilde{E}_{i}^{m, \, n \, (\ell \, + )} 
 = \left(
\begin{array}{c} 
\ell \\
n 
\end{array} 
 \right) \, 
\left[
\begin{array}{c} 
\ell \\
m 
\end{array} 
 \right]_q \, 
\left[
\begin{array}{c} 
\ell \\
n 
\end{array} 
 \right]_q^{-1} \, 
\widetilde{P}^{(\ell)}_{1 \cdots L} 
\,  
\prod_{\alpha=1}^{(i-1) \ell} (A^{(1+)}+D^{(1+)}(w_{\alpha}) 
\nonumber \\ 
& &  \times
\prod_{k=1}^{m} D^{(1+)}(w_{(i-1)\ell+k} ) 
\prod_{k=m+1}^{n} C^{(1+)}(w_{(i-1)\ell + k})  
\, \prod_{k=n+1}^{\ell} A^{(1+)}(w_{(i-1) \ell+k})  
\nonumber \\ 
& &  \times
\prod_{\alpha=i \ell +1}^{\ell N_s} 
(A^{(1+)}+D^{(1+)})(w_{\alpha}) \, \,   
\widetilde{P}^{(\ell)}_{1 \cdots L}
\, .  \label{eq:Em<n}
\end{eqnarray}

By the quantum inverse-scattering problem (QISP)
of Ref. \refcite{MT2000} the local spin operators are expressed in terms of global operators and the transfer matrices for the integrable spin-$s$ 
XXX spin chain. However, it is not clear how one can derive 
(\ref{eq:Em>n}) and (\ref{eq:Em<n}) 
by the QISP method even for $q=1$. 

%
%
\subsection{Symbols for expressing sequences}

Let us denote by $(a_j)_m$
a sequence of numbers $a_j$ for $j=1, 2, \ldots, m$, i.e.     
$(a_j)_m = (a_1, a_2, \cdots, a_m)$. 

\begin{definition}
We say that a sequence $(b_k)_n$ is a subsequence of $(a_j)_m$ if 
(i) $n \le m$, (ii) $b_k \in \{a_1, \ldots, a_m\}$ 
for $k=1, 2, \ldots, n$, (iii) for any  pair of integers $j$ and $k$ 
satisfying $1 \le j <  k \le n$, there exists a pair of integers 
$\ell(j)$ and $\ell(k)$ such that 
$a_j = b_{\ell(j)}$, $a_k = b_{\ell(k)}$ and  $\ell(j) < \ell(k)$.   
\end{definition}

For a pair of sequences $(a_j)_m$ and $(b_k)_n$,  we define 
the product $(a_j)_m \# (b_k)_n$ by a sequence $(c_{\ell})_{m+n}$ 
such that $c_j = a_j $ for $j=1, 2, \cdots, m$ and  
$c_j= b_j$ for $j=m+1, m+2, \ldots, m+n$.

%
%
\subsection{Conjecture of the spin-$s$ Ground-state solution}

Let us now introduce the conjecture that the ground state 
of the spin-$s$ case $| \psi_g^{(2s \, +)} \rangle$ is 
given by $N_s/2$ sets of $2s$-strings:   
\begin{equation} 
\lambda_{a}^{(\alpha)} 
= \mu_a - (\alpha- 1/2) \eta + \epsilon_a^{(\alpha)} \, , \quad  
\mbox{for} \, \, a=1, 2, \ldots, N_s/2 \, \,  
\mbox{and} \, \,  \alpha = 1, 2, \ldots, 2s .  
\end{equation} 
Here we assume that string deviations $\epsilon_a^{(\alpha)}$ are  
very small when $N_s$ is very large \cite{Takahashi-book}. 
In terms of $\lambda_{a}^{(\alpha)}$, 
the spin-$s$ ground state in the homogeneous grading 
is given by \cite{DM2}  
\begin{equation}  
 | \psi_g^{(2s \, +)} \rangle = 
\prod_{a=1}^{N_s/2} \prod_{\alpha=1}^{2s} 
\widetilde{B}^{(2s \, +)}(\lambda_a^{(\alpha)}; \{\xi_p \}_{N_s}) 
| 0 \rangle . 
\end{equation}
We denote by $M$ the number of Bethe roots:  
$M= 2s \, N_s/2 = s N_s$.

According to analytic and numerical studies 
\cite{deVega-Woynarovich,KR,KBP,JSuzuki},  
we may assume the following  
properties of string deviations $\epsilon_a^{(\alpha)}$s. 
For very large $N_s$, the deviations are given by  
$\epsilon_a^{(\alpha)} = i  \, \delta_a^{(\alpha)}$, 
where $i$ denotes $\sqrt{-1}$ and  
$\delta_a^{(\alpha)}$ are real. Moreover, 
$\delta_{a}^{(\alpha)} - \delta_{a}^{(\alpha+1)} > 0$ 
for $\alpha=1, 2, \ldots, 2s-1$,  and 
$|\delta_{a}^{(\alpha)}| > |\delta_{a} ^{(\alpha+1)}|$ 
for $\alpha < s$, 
 while $|\delta_{a}^{(\alpha)}| < |\delta_{a} ^{(\alpha+1)}|$ 
for $\alpha \ge s$.

In the limit: $N_s \rightarrow \infty$,  
the density of string centers, $\rho_{\rm tot}(\mu)$, is given by 
\begin{equation} 
\rho_{\rm tot}(\mu)= {\frac 1 N_s} \sum_{p=1}^{N_s} 
{\frac 1 {2 \zeta \cosh(\pi (\mu- \xi_p)/\zeta)}} \, . 
\end{equation}
For the homogeneous chain where $\xi_p=0$ for $p=1, 2, \ldots, N_s$, 
we denote the density of string centers by $\rho(\lambda)$.  
\begin{equation}  
\rho(\lambda)= {\frac 1 {2 \zeta \cosh(\pi \lambda/ \zeta)}} \, . 
\end{equation}

Let us introduce useful notation of the suffix of rapidities. 
For rapidities $\lambda_{a}^{(\alpha)}=\lambda_{(a, \alpha)}$ 
we define integers $A$ by $A= 2s(a-1) + \alpha$ for 
$a=1, 2, \ldots, N_s/2$ and for $\alpha=1, 2, \ldots, 2s$.  
We thus denote $\lambda_{(a, \alpha)}$ also 
by $\lambda_A$ for $A=1, 2, \ldots, s N_s$, and 
put $\lambda_{(a, \alpha)}$ 
in increasing order with respect to $A=2s(a-1)+\alpha$ 
such as $\lambda_{(1,1)}=\lambda_1, \lambda_{(1,2)}=\lambda_2, 
\ldots, \lambda_{(N_s/2, 2s)}=\lambda_{s N_s}$.    
In the ground state,  
 rapidities $\lambda_A$ for 
$A=1, 2, \ldots, M$, are expressed by   
\begin{equation}  
\lambda_{2s(a-1)+ \alpha} = \mu_{a} - (\alpha- 1/2) \eta 
+ \epsilon_a^{(\alpha)} \quad
\, (1 \le a \le N_s/2; \, \, 1 \le \alpha \le 2s).   
\label{eq:2s-string}
\end{equation} 
For $A=2s(a-1)+ \alpha$ with $1 \le \alpha \le 2s$, 
integer $a$ is given by $a=[(A-1)/2s]+1$, and integer $\alpha$ 
is given by $\alpha=A - 2s [(A-1)/2s]$. 

For a real number $x$ we define $[x]$  
by the greatest integer less than or equal to $x$. 
We define $a(j)$ and $\alpha(j)$ for $j=1, 2, \ldots, M$ as follows.  
\begin{equation} 
a(j)=[(j-1)/2s]+1 \, , \quad \alpha(j)=j - 2s [(j-1)/2s] \, . 
\label{eq:a(j)}
\end{equation}

%
%
\subsection{Correlation functions 
of the integrable spin-$s$ XXZ model on a long finite chain}

We define the correlation function of the integrable spin-$2s$ XXZ spin chain 
for a given product of $(2s+1) \times (2s+1)$ elementary matrices 
such as $\widetilde{E}_1^{i_1 , \, j_1 \, (2s \, +)} \cdots 
\widetilde{E}_m^{i_m, \, j_m \, (2s \, +)}$ 
on the spin-$s$ ground state, 
$| \psi_g^{(2s \, +)} \rangle$, 
as follows. 
\begin{equation}
F_m^{(2s \, +)}(\{i_k, j_k \}) =  \langle \psi_g^{(2s \, +)} | 
\prod_{k=1}^{m} \widetilde{E}_k^{i_k , \, j_k \, (2s \, +)} 
|\psi_g^{(2s \, +)}  \rangle / \langle \psi_g^{(2s \, +)} 
| \psi_g^{(2s \, +)} \rangle \, . 
\label{eq:def-CF}
\end{equation}
By formulas (\ref{eq:Em>n}) and (\ref{eq:Em<n}) 
we express the $m$th product of $(2s+1) \times (2s+1)$ elementary matrices  
in terms of a $2sm$th product of $2 \times 2$ elementary matrices  
with entries $\{ \epsilon_j, \epsilon_j^{'} \}$ as follows. 
\begin{equation} 
\prod_{b=1}^{m} \widetilde{E_b}^{i_b , \, j_b \, (2s \, +)} = 
C(\{i_b, j_b \}) 
\, 
\widetilde{P}_{1 2 \ldots L}^{(2s)} \, \cdot \,  
\prod_{k=1}^{2sm} 
 e_k^{\epsilon_k^{'}, \,  \epsilon_k} \, \cdot \, 
\widetilde{P}_{1 2 \ldots L}^{(2s)} . 
\end{equation}
By making use of (\ref{eq:Em>n}) and (\ref{eq:Em<n}),  
$C(\{i_b, j_b \})$ is given by 
\begin{equation} 
C(\{i_k, j_k \}) = \prod_{b=1}^{m} \left\{  
\left(\begin{array}{c}
      2s \\
      j_b  
      \end{array} 
 \right) 
\left[ \begin{array}{c}
      2s \\
      i_b  
      \end{array} 
 \right]_q  
\left[ \begin{array}{c}
      2s \\
      j_b  
      \end{array} 
 \right]_q^{-1}  
\right\} \, .  
\end{equation}
Here $\epsilon_{2s(b-1)+ \beta}$ and $\epsilon_{2s(b-1)+ \beta}^{'}$ 
($b=1, \ldots, N_s$; $\beta=1, \ldots, 2s$) are given by  
\begin{equation}
\epsilon_{2s(b-1)+ \beta} =  
\left\{ \begin{array}{ccc} 
        1  &  & (1 \le \beta \le j_b) \\ 
        0  &  & (j_b < \beta \le 2s)  
         \end{array} \right.   \, ; \, \,   
\epsilon_{2s(b-1)+ \beta}^{'} = 
\left\{ \begin{array}{ccc} 
        1  &  & (1 \le \beta \le i_b) \\ 
        0  &  & (i_b < \beta \le 2s) \, . 
         \end{array} \right. 
 \label{eq:epsilon}
\end{equation} 
We evaluate the spin-$2s$ XXZ correlation function 
 $F_m^{(2s \, +)}(\{i_k, j_k \}) $ by 
\begin{eqnarray}
F_m^{(2s \, +)}(\{i_k, j_k \}) & = & C(\{i_k, j_k\}) \, 
\langle \psi_g^{(2s+)} | \,  \widetilde{P}_{1 2 \ldots L}^{(2s)} 
\, \, \times \, \nonumber \\ 
& &  \times \, 
\prod_{j=1}^{2sm} e_j^{\epsilon_j^{'}, \, \epsilon_j} 
 \, \cdot \, \widetilde{P}_{1 2 \ldots L}^{(2s)} 
|\psi_g^{(2s+)}  \rangle / \langle \psi_g^{(2s+)} | \psi_g^{(2s+)}  
\rangle \, . 
\label{eq:CFe}
\end{eqnarray}

Let $\mbox{\boldmath$\alpha$}^{+}$ be 
the set of  $j$ with $\epsilon_j=0$, 
and $\mbox{\boldmath$\alpha$}^{-}$ 
the set of $j$ with $\epsilon_j^{'}=1$:  
\begin{equation} 
\mbox{\boldmath$\alpha$}^{+}= \{ j ; \, \epsilon_j=0 \}  \, , \quad 
\mbox{\boldmath$\alpha$}^{-}= \{ j ; \, \epsilon_j^{'}=1 \} \,. 
\label{eq:def-aa'}
\end{equation} 
We denote by $r$ and $r^{'}$ the number 
of elements of the set $\mbox{\boldmath$\alpha$}^{-}$ 
and $\mbox{\boldmath$\alpha$}^{+}$, respectively. Due to charge conservation, 
we have 
\begin{equation} 
r + r^{'} = 2sm .  
\end{equation}
We denote by $j_{\rm min}$ and $j_{\rm max}$ the smallest element and 
the largest element of $\mbox{\boldmath$\alpha$}^{-}$, respectively.    
We also denote by $j^{'}_{\rm min}$ and $j^{'}_{\rm max}$ 
the smallest element and the largest element 
of $\mbox{\boldmath$\alpha$}^{+}$, respectively.  

Recall that the ground state $| \psi_g^{(2s \, +)} \rangle$ 
has $M$ Bethe roots with $M=sN_s$.  
Let  $c_j$ ($j \in \mbox{\boldmath$\alpha$}^{-}$) and 
 $c_j^{'}$ ($j \in \mbox{\boldmath$\alpha$}^{+}$)
be integers such that 
 $1 \le c_j \le M$ for $j \in \mbox{\boldmath$\alpha$}^{-}$ 
and $1 \le c_j^{'} \le M+j$ for $j \in \mbox{\boldmath$\alpha$}^{+}$.  
We define sequence $(b_{\ell})_{2sm}$ by 
\begin{equation} 
(b_1 , b_2, \ldots, b_{2sm}) = 
(c_{j_{\rm max}^{'}}^{'}, \ldots, c^{'}_{j_{\rm min}^{'}}, 
c_{j_{\rm min}}, \ldots, 
c_{j_{\rm max}}) \, .  \label{df:seq-b}  
\end{equation}
Here sequence 
$(c_{j_{\rm max}^{'}}^{'}, \ldots, c^{'}_{j_{\rm min}^{'}}, 
c_{j_{\rm min}}, \ldots, c_{j_{\rm max}})$ is given by the 
composite sequence of 
 $c^{'}_{j}$ s in decreasing order with respect to suffix $j$, 
and $c_{j}$ s in increasing order with respect to suffix $j$.    
We introduce the following symbols: 
\begin{equation} 
\prod_{j \in \mbox{\boldmath$\alpha$}^{-}} 
\left( \sum_{c_j=1}^{M} \right)   
\prod_{j \in \mbox{\boldmath$\alpha$}^{+}} 
\left( \sum_{c_j^{'}=1}^{M+j} \right) 
=
 \sum_{c_{j_{\rm min}}=1}^{M} \cdots \sum_{c_{j_{\rm max}}=1}^{M} 
\sum_{c^{'}_{j^{'}_{\rm min}}=1}^{M+j^{'}_{\rm min}} \cdots 
\sum_{c^{'}_{j^{'}_{\rm max}}=1}^{M+j^{'}_{\rm max}} \,. 
\label{eq:symbol}
\end{equation}    
Recall that $a(j)$ are defined in (\ref{eq:a(j)}). 
We define $\beta(j)$ by 
\begin{equation}
\beta(j) = j - 2s [(j-1)/2s] \quad (1 \le j \le M). \label{df:beta}
\end{equation}   
For $\ell, k = 1, 2, \ldots, 2sm$, we define 
the $(\ell, k)$ element  of $M^{(2sm)}((b_{j})_{2sm})$ by 
\begin{eqnarray} 
& & \left( M^{(2sm)}((b_{j})_{2sm}) \right)_{\ell, \,  k} 
\nonumber \\ 
& &  = \left\{ 
\begin{array}{cc}  
- \delta_{b_{\ell}-M, \, k} &  \, (b_{\ell} > M)  \\ 
\delta_{\beta(b_{\ell}), \, \beta(k)} \cdot 
\rho(\lambda_{b_{\ell}}- w_k^{(2s)} + \eta/2 ) 
/({N_s \rho_{\rm tot}(\mu_{a(b_{\ell})})}) 
 &   \, (b_{\ell} \le M)  
\end{array}
\right.   
\end{eqnarray}
Here, continuous variable $\mu$, which is the argument of 
density $\rho_{\rm tot}(\mu)$,  is evaluated at 
$\mu_{a(b_{\ell})}$, one of the ``string centers'' $\mu_a$  
of $2s$-strings (\ref{eq:2s-string}).

We can rigorously derive a concise expression of  
correlation functions of the spin-$s$ XXZ spin chain 
in the massless region: $0 \le \zeta < \pi/2s$ for a large finite chain.  
Introducing $\varphi(\lambda)= \sinh \lambda$ we have    
\begin{eqnarray} 
& & F_{m}^{(2s \, +)}(\{i_b, j_b \}) 
 =  C(\{i_k, j_k\})  \prod_{j \in \mbox{\boldmath$\alpha$}^{-}} 
\left( \sum_{c_j=1}^{M} \right)   
\prod_{j \in \mbox{\boldmath$\alpha$}^{+}} 
\left( \sum_{c_j^{'}=1}^{M+j} \right) 
 {\rm det} M^{(2sm)}((b_{\ell})_{2sm})  
\nonumber 
\\ 
& & \, \, \times  
(-1)^{r^{'}} 
{\frac { \prod_{j \in \mbox{\boldmath$\alpha$}^{-}} 
\left( \prod_{k=1}^{j-1} 
\varphi(\lambda_{c_j} - w_k^{(2s)} + \eta) 
\prod_{k=j+1}^{2sm} \varphi(\lambda_{c_j} - w_k^{(2s)} ) \right)}
{\prod_{1 \le k < \ell \le 2sm} 
\varphi(\lambda_{b_{\ell}} - \lambda_{b_k} + \eta)} } \nonumber 
\end{eqnarray}
\begin{eqnarray}
& & \, \, \times  
{\frac { \prod_{j \in \mbox{\boldmath$\alpha$}^{+}} 
\left( \prod_{k=1}^{j-1} 
\varphi(\lambda_{c_j^{'}} - w_k^{(2s)} - \eta) 
\prod_{k=j+1}^{2sm} \varphi(\lambda_{c_j^{'}} - w_k^{(2s)} ) \right)}
{\prod_{1 \le k < \ell \le 2sm} 
\varphi(w_{k}^{(2s)} - w_{\ell}^{(2s)})} } + O(1/N_s) \, . 
\nonumber \\ 
\label{eq:finiteCF}
\end{eqnarray}
We remark that we derive 
(\ref{eq:finiteCF}) sending $\epsilon$ to zero. 
Before taking the limit, 
inhomogeneous parameters $w_j$s are generic 
due to small parameter $\epsilon$, 
and the sums over variables $c_j$ in (\ref{eq:finiteCF}) 
are restricted up to $M$ for all $j$.

%
%
\subsection{Multiple-integral representations of spin-$s$ XXZ correlation function for arbitrary matrix elements}

In the thermodynamic limit: $N_s \rightarrow \infty$, 
rapidities $\lambda_{b_{\ell}}$ with 
$b_{\ell}$ defined in (\ref{df:seq-b}),  
correspond to integral variables $\lambda_{\ell}$ for 
$\ell =1, 2, \ldots, 2sm$.  For $1 \le b_{\ell} \le M$ 
they are given by the Bethe roots of $2s$-strings (\ref{eq:2s-string}), 
while for $b_{\ell} > M$ they are given by 
complete $2s$-strings $w_{k}^{(2s)}$ defined by (\ref{eq:2s-strings}).

We define $\alpha(\lambda_j)$ by $\alpha(\lambda_j)= \gamma$ 
for an integer $\gamma$ with $1 \le \gamma \le 2s$,  
if $\lambda_j$ is related to integral variable $\mu_j$ 
by $\lambda_j = \mu_j - (\gamma - 1/2) \eta$, 
or if $\lambda_j$ takes a value close to $w_k^{(2s)}$ with $\beta(k)=\gamma$, 
where $w_k^{(2s)}$ are part of complete 2s-strings (\ref{eq:2s-strings}). 
Here, variables $\mu_j$ correspond to ``string centers'' of 
variables $\lambda_j$.

We define the $(j,k)$ element of matrix 
$S=S\left( (\lambda_j)_{2sm}; (w_j^{(2s)})_{2sm} \right)$ by   
\begin{equation} 
S_{j,k} = \rho(\lambda_j - w_k^{(2s)} + \eta/2) \, 
\delta(\alpha(\lambda_j), \beta(k)) \, , \quad {\rm for} \quad 
j, k= 1, 2, \ldots, 2sm \, .  
\end{equation} 
Here $\delta(\alpha, \beta)$ denotes the Kronecker delta, and 
we recall (\ref{df:beta}) for $\beta(k)$.

Let  $\Gamma_j$ be a small contour rotating counterclockwise   
around $\lambda=w_j^{(2s)}$.   
Since ${\rm det} S$ has simple poles at $\lambda=w_j^{(2s)}$ 
with residue $1 / 2 \pi i$,  
we have 
\begin{equation} \int_{-\infty+ i \epsilon}^{\infty+ i \epsilon} 
{\rm det} S((\lambda_k)_{2sm}) \, 
 d \lambda_1 = 
\int_{-\infty - i \epsilon}^{\infty - i \epsilon} 
{\rm det} S((\lambda_k)_{2sm}) \, d \lambda_1 
- \oint_{\Gamma_1} {\rm det} S((\lambda_k)_{2sm}) 
\, d \lambda_1 \, .  
\end{equation}
For sets  ${\bm \alpha}^{-}$ and ${\bm \alpha}^{+}$ with 
relation (\ref{df:seq-b}), 
we define integral variables ${\tilde \lambda}_j$ 
for $j \in {\bm \alpha}^{-}$ 
and ${\tilde \lambda}^{'}_{j}$ for $j \in {\bm \alpha}^{+}$, 
respectively,  by the following:   
\begin{equation} 
({\tilde \lambda}^{'}_{j^{'}_{max}}, \ldots, 
{\tilde \lambda}^{'}_{j^{'}_{min}},  {\tilde \lambda}_{j_{min}}, 
{\tilde \lambda}_{j_{max}})
=(\lambda_1, \ldots, \lambda_{2sm}) \, . 
\end{equation}
Thus, from expression (\ref{eq:finiteCF}) of  
the correlation function in terms of a finite sum,   
we derive the multiple-integral representation as follows.     
\begin{eqnarray} 
& & F^{(2s+)}_{m}(\{i_k, j_k\}) =  
\quad C(\{i_b, j_b \}) \, \times 
\nonumber \\ 
& & 
\times \left( \int_{-\infty+ i \epsilon}^{\infty+ i \epsilon}
+ \cdots 
+ \int_{-\infty - i \widetilde{ \zeta_s} 
+ i \epsilon}
^{\infty - i \widetilde{\zeta_s} 
+ i \epsilon} \right)  d \lambda_1 
\cdots 
\left( \int_{-\infty+ i \epsilon}^{\infty+ i \epsilon}
+ \cdots 
+ \int_{-\infty - i \widetilde{\zeta_s } 
+ i \epsilon}^{\infty - i \widetilde{ \zeta_s }
 + i \epsilon} 
\right)  d \lambda_{r^{'}} 
\nonumber \\  
& & \times \left( \int_{-\infty - i \epsilon}^{\infty - i \epsilon}
+ \cdots 
+ \int_{-\infty - i \widetilde{\zeta_s} 
- i \epsilon}
^{\infty - i \widetilde{\zeta_s} 
 - i \epsilon} \right)  d \lambda_{\widetilde{r}}
\cdots 
\left( \int_{-\infty - i \epsilon}^{\infty - i \epsilon}
+ \cdots 
+ \int_{-\infty - i \widetilde{\zeta_s} 
 - i \epsilon}
^{\infty - i \widetilde{\zeta_s} 
 - i \epsilon} 
\right)  d \lambda_{2sm}
\nonumber \\ 
& & \quad \times Q(\{ \epsilon_j, \epsilon_j^{'} \}; \lambda_1, \ldots, \lambda_{2sm}) \, {\rm det}
S(\lambda_1, \ldots, \lambda_{2sm}) \, . 
\label{eq:MIR}
\end{eqnarray}
Here $\widetilde{\zeta_s} = (2s-1) \zeta$, 
$\widetilde{ r}=r^{'} + 1$, and  
$Q(\{ \epsilon_j, \epsilon_j^{'} \}; \lambda_1, \ldots, \lambda_{2sm})$ 
is given by   
\begin{eqnarray} 
& & Q(\{ \epsilon_j, \epsilon_j^{'} \}; \lambda_1, \ldots, \lambda_{2sm}) 
\nonumber \\ 
& & 
=(-1)^{r^{'}} 
{\frac { \prod_{j \in {\bm \alpha}^{-}} \left( \prod_{k=1}^{j-1} 
\varphi({\tilde \lambda}_{j} - w_k^{(2s)} + \eta) 
\prod_{k=j+1}^{2sm} \varphi({\tilde \lambda}_{j} - w_k^{(2s)} ) \right)}
{\prod_{1 \le k < \ell \le 2sm} 
\varphi(\lambda_{\ell} - \lambda_{k} + \eta + \epsilon_{\ell, k})} } \nonumber \\ 
& \times &  
{\frac { \prod_{j \in {\bm \alpha}^{+}} \left( \prod_{k=1}^{j-1} 
\varphi({\tilde \lambda}^{'}_{j} - w_k^{(2s)} - \eta) 
\prod_{k=j+1}^{2sm} \varphi({\tilde \lambda}^{'}_{j} - w_k^{(2s)} ) \right)}
{\prod_{1 \le k < \ell \le 2sm} 
\varphi(w_{k}^{(2s)} - w_{\ell}^{(2s)})} } \, . 
\end{eqnarray}
In the denominator we set $\epsilon_{k, \ell}= i \epsilon$ for 
${Im}(\lambda_k -\lambda_{\ell}) > 0$ and 
$\epsilon_{k, \ell}= - i \epsilon$ for 
${Im}(\lambda_k -\lambda_{\ell}) < 0$,  where
$\epsilon$ is infinitesimally small: $|\epsilon| \ll 1$.  Here, 
${Im} (a+ i b) = b$ for real numbers $a$ and $b$.   
Here, for ${\bm \alpha}^{\pm}$,  we recall (\ref{eq:def-aa'}). 

We evaluate $\alpha(\lambda_j)$ in (\ref{eq:MIR}), replacing 
paths $(- \infty - i (\gamma-1) \zeta \pm i \epsilon, 
\infty - i (\gamma-1) \zeta \pm i \epsilon)$ 
by $(- \infty - i (\gamma-1/2)  \zeta, 
\infty - i (\gamma-1/2) \zeta)$ for $ \gamma=1, 2, \ldots, 2s$, 
respectively. The integrals over $\lambda_j$ for $j \ge {\tilde r}$ 
do not change when $\epsilon \rightarrow \zeta/2$. 

Thus, correlation functions (\ref{eq:def-CF}) are expressed 
in the form of a single term of multiple integrals (\ref{eq:MIR}).

We can derive the symmetric expression for the multiple-integral 
representations of the spin-$s$ correlation function 
$F^{(2s+)}_{m}(\{i_k, j_k\})$ as follows \cite{DM2}. 
\begin{eqnarray}
& & F^{(2s+)}_{m}(\{i_k, j_k\}) =  
C(\{i_b, j_b \}) \times 
\nonumber \\ 
&& \begin{split}
& 
\times  \frac{1}{\prod_{1 \leq \alpha < \beta \leq 2s}
\sinh^{m}(\beta-\alpha )\eta}     
\prod_{1\leq k < l \leq m}
\frac{\sinh^{2s}(\pi(\xi_k-\xi_l)/\zeta)}
{\prod^{2s}_{j=1}\prod^{2s}_{r=1}\sinh(\xi_k-\xi_l+(r-j)\eta)}   
 \\
& \times \sum_{\sigma \in {\cal S}_{2sm}/({\cal S}_m)^{2s} } 
({\rm sgn} \, \sigma) \, 
 \prod^{r^{'}}_{j=1} 
\left( 
\int^{\infty+ i \epsilon }_{-\infty + i \epsilon} + 
\cdots + 
\int^{\infty - i(2s-1) \zeta + i \epsilon }_{-\infty  - i(2s-1) \zeta  
+ i  \epsilon} 
\right) 
d \mu_{\sigma j} \\  
& \times 
\prod^{2sm}_{j=r^{'} +1}
\left( 
\int^{\infty - i \epsilon }_{-\infty - i \epsilon} + 
\cdots 
+ \int^{\infty - i(2s-1) \zeta - i \epsilon }_{-\infty  - i(2s-1) \zeta  
- i  \epsilon} 
\right) d \mu_{\sigma j} 
\nonumber \\  
&
\times 
Q(\{ \epsilon_j, \epsilon_j^{'} \}; \lambda_{\sigma 1}, \ldots, 
\lambda_{\sigma(2sm)})) \, 
\left( \prod^{2sm}_{j=1} 
{\frac {\prod^{m}_{b=1} \prod^{2s-1}_{\beta=1}
\sinh(\lambda_{j}-\xi_b+ \beta \eta)}
{\prod_{b=1}^{m} \cosh(\pi(\mu_{j}-\xi_b)/\zeta)}} \right) \\ 
&\times \, {\frac {i^{2sm^2}} { (2 i \zeta)^{2sm} }} \, 
\prod^{2s}_{\gamma=1} 
\prod_{1 \le b < a \le m}
\sinh(\pi(\mu_{2s(a-1)+\gamma}-\mu_{2s(b-1)+\gamma})/\zeta) \, . 
\end{split} 
\nonumber \\ 
 \label{eq:CFF2}
\end{eqnarray}
Here $\lambda_j$ are given by $\lambda_j= \mu_j - (\beta(j)-1/2) \eta$ for 
$j=1, \ldots, 2sm$.  

It is straightforward to take the homogeneous limit: 
$\xi_k \rightarrow 0$. 
Here  (${\rm sgn} \, \sigma$) 
denotes the sign of permutation 
$\sigma \in {\cal S}_{2sm}/({\cal S}_m)^{2s}$.

%
%
 \setcounter{equation}{0} 
 \renewcommand{\theequation}{6.\arabic{equation}}
\section{Derivation of finite-sum expression of 
spin-$s$ XXZ correlation functions with arbitrary entries}

\subsection{Fundamental commutation relations}

We now discuss briefly the derivation of (\ref{eq:finiteCF}), 
which expresses the spin-$s$ XXZ correlation functions with arbitrary entries 
in terms of the product of finite sums over the Bethe roots.

Let $\Sigma_N$ be the set of integers $1, 2, \ldots, N$, i.e. 
$\Sigma_N= \{ 1, 2, \ldots, N \}$. Recall 
definition (\ref{eq:def-aa'}) of $\mbox{\boldmath$\alpha$}^{\pm}$ and 
that of integers $c_j$ and $c_j^{'}$. For a given set of $c_j, c_j^{'}$, 
we introduce $\mbox{\boldmath$A$}_j$ and $\mbox{\boldmath$A$}_j^{'}$ by 
\begin{eqnarray} 
\mbox{\boldmath$A$}_j & = & 
\{ b; 1 \le b \le M + 2sm, \, b \ne c_k, c_k^{'} \, \, 
\mbox{for} \, k < j \} \, , 
\nonumber \\ 
\mbox{\boldmath$A$}_j^{'} 
& = & \{ b; 1 \le b \le M+2sm, \, b \ne c_k \, \, \mbox{for} \, k \le j, 
b \ne c_k^{'} \, \mbox{for} \, k < j \}.          
\end{eqnarray}
We define sets $\mbox{\boldmath$\alpha$}_j^{\pm}$ and 
$c(\mbox{\boldmath$\alpha$}_j^{\pm})$ as follows.    
\begin{eqnarray}
& & \mbox{\boldmath$\alpha$}_j^{-} = 
\{k ; \, k < j, k \in \mbox{\boldmath$\alpha$}^{-} \} \, , \quad  
\mbox{\boldmath$\alpha$}_j^{+} = \{k ; \, k < j, k 
\in \mbox{\boldmath$\alpha$}^{+} \} \, , \\  
& & c(\mbox{\boldmath$\alpha$}_j^{-}) = 
\{c_k; \, k \in \mbox{\boldmath$\alpha$}_j^{-}\} \, , \quad      
c(\mbox{\boldmath$\alpha$}_j^{+})=
\{c_k^{'}; \, k \in \mbox{\boldmath$\alpha$}_j^{+}\} \, . 
\end{eqnarray}    
We have  
\begin{equation} 
\mbox{\boldmath$A$}_j = \Sigma_{M+2sm} \setminus 
\left(  c(\mbox{\boldmath$\alpha$}_j^{-}) \cup c(\mbox{\boldmath$\alpha$}_j^{+}) \right) \, , \,  
\mbox{\boldmath$A$}_j^{'} = \Sigma_{M+2sm} \setminus 
\left(  c(\mbox{\boldmath$\alpha$}_{j+1}^{-}) \cup c(\mbox{\boldmath$\alpha$}_j^{+}) \right) \, . \nonumber 
\end{equation}

Let us denote by $t$ the number of $c_j$ 
($j \in \mbox{\boldmath$\alpha$}^{-}$) and   
 $c_j^{'}$  ($j \in \mbox{\boldmath$\alpha$}^{+}$)  
such that $c_j, c_j^{'} \le M$, for a given set of 
$c_j$ and $c_j^{'}$.  
We express (\ref{eq:symbol}) as follows. 
\begin{equation} 
\sum_{t=r}^{2sm} \sum_{\{ c_j, c_j^{'} \}_t} 
= \prod_{j \in \mbox{\boldmath$\alpha$}^{-}} 
\left( \sum_{c_j=1}^{M} \right)   
\prod_{j \in \mbox{\boldmath$\alpha$}^{+}} 
\left( \sum_{c_j^{'}=1}^{M+j} \right) \, . 
\end{equation}
Here the sum over ${\{ c_j, c_j^{'} \}_t}$ denotes the sums 
over $c_j$ and $c_j^{'}$ such that the number of 
$c_j^{'} \le M$ is fixed by $t-r$. 

Suppose that 
$\lambda_{\alpha}$ for $\alpha=1, 2, \ldots, M$ give a set of 
solutions of the Bethe ansatz equations in the spin-1/2 case with 
$w_j =w_j^{(2s; \, \epsilon)}$ for $j=1, 2, \ldots, L$. \cite{DM2}
Here $w_j$ are inhomogeneous parameters. 
We set rapidities $\lambda_{M+j}$  by 
\begin{equation} 
\lambda_{M+j}= w_j \, ,  \qquad j=1, 2, \ldots, 2sm \, .  
\end{equation} 
We can show the fundamental commutation relations as follows \cite{KMT2000}.   
\begin{eqnarray}  
& & \langle 0 | \left( \prod_{\alpha=1}^{M} C(\lambda_{\alpha}) \right) \, 
T_{\epsilon_1, \epsilon_1^{'}}(\lambda_{M+1}) \cdots  
T_{\epsilon_{2sm}, \epsilon_{2sm}^{'}}(\lambda_{M+2sm})  
\nonumber \\ 
& = & 
\sum_{t=r}^{2sm} \sum_{\{ c_j, c_j^{'} \}_t }  
G_{\{ c_j, \,  c_j^{'} \}}(\lambda_1, \cdots, \lambda_{M+2sm})      
\langle 0 | 
\prod_{k \in \mbox{\boldmath$A$}_{2sm+1}(\{c_j, c_j^{'} \}) } 
C(\lambda_{k}) \, ,  \nonumber   
\end{eqnarray}
where $d(\lambda; \{ w_k^{(2s; \epsilon)} \}_L)$ and 
$G_{\{ c_j, c_j^{'} \}}((\lambda_{\alpha})_{M+2sm})$ are given by   
\begin{eqnarray} 
& & d(\lambda; \{ w_k^{(2s; \epsilon)} \}_L) 
 =  \prod_{k=1}^{L} b(\lambda- w_k^{(2s; \epsilon)}) \, , 
\nonumber \\  
& & G_{\{ c_j, c_j^{'} \}}(\lambda_1, \cdots, \lambda_{M+2sm})  
= \prod_{j \in \mbox{\boldmath$\alpha$}^{+} }
 \left( 
{\frac {\prod_{b=1; b \in \mbox{\boldmath$A$}^{'}_j }^{M+j-1} 
\varphi(\lambda_{b}-\lambda_{c_j^{'}} + \eta) }  
 {\prod_{b=1, b \in \mbox{\boldmath$A$}_{j+1} }^{M+j} 
 \varphi(\lambda_{b} - \lambda_{c_j^{'}})}}  
\right) 
\nonumber \\ 
& & \qquad \qquad  
\times  
 \prod_{j \in \mbox{\boldmath$\alpha$}^{-} }
 \left( d(\lambda_{c_j}; \{ w_k^{(2s; \, \epsilon)} \}_L) 
{\frac {\prod_{b=1; b \in \mbox{\boldmath$A$}_j}^{M+j-1} 
\varphi(\lambda_{c_j}-\lambda_b + \eta) }  
 {\prod_{b=1, b \in \mbox{\boldmath$A$}^{'}_j }^{M+j} 
 \varphi(\lambda_{c_j} - \lambda_b)}}  
     \right)
 \, . \nonumber \\
\label{eq:FCR}
\end{eqnarray}

%
%
\subsection{Finite-sum expression of correlation functions for a finite chain}

We introduce disjoint subsets of $\mbox{\boldmath$\alpha$}^{+}$,  
$\mbox{\boldmath$\alpha$}_J^{+}$ and $\mbox{\boldmath$\alpha$}_K^{+}$, as follows.   
\begin{equation} 
\mbox{\boldmath$\alpha$}_J^{+} = \{ j ; \, j \in \mbox{\boldmath$\alpha$}^{+}, 1 \le c_j^{'} \le M \} \, ,    
\quad 
\mbox{\boldmath$\alpha$}_K^{+} = \{ j ; j \, \in \mbox{\boldmath$\alpha$}^{+}, c_j^{'} > M \} \, .  
\end{equation} 
We define sets 
$c(\mbox{\boldmath$\alpha$}^{-})$, 
$c(\mbox{\boldmath$\alpha$}^{+}_J)$ and 
$c(\mbox{\boldmath$\alpha$}^{+}_K)$ as follows.  
\begin{equation}  
c(\mbox{\boldmath$\alpha$}^{-})=
\{c_k; \, k \in \mbox{\boldmath$\alpha$}^{-}\} , \, 
c(\mbox{\boldmath$\alpha$}_J^{+})=
\{c_k; \, k \in \mbox{\boldmath$\alpha$}_J^{+}\} , \,  
c(\mbox{\boldmath$\alpha$}_K^{+})=
\{c_k; \, k \in \mbox{\boldmath$\alpha$}_K^{+}\} \, .  
\nonumber 
\end{equation}
We define a sequence $(\widetilde{b_k})_t$ by a subsequence of 
$(b_k)_{2sm}$ such that ${\tilde b}_k \le M$ for $k=1, 2, \ldots, t$.   
We denote sequence $(b_k)_{2sm}$ and $(\widetilde{b_k})_t$ as sets 
by $\mbox{\boldmath$b$}$ and $\widetilde{\mbox{\boldmath$b$}_t}$, 
respectively, i.e. 
$\mbox{\boldmath$b$}=\{ b_1, b_2, \ldots, b_{2sm} \}$ and    
$\widetilde{\mbox{\boldmath$b$}_t}=\{ \widetilde{b}_1, \ldots, 
\widetilde{b}_t \}$. 
Here we note 
$\widetilde{\mbox{\boldmath$b$}_t}=
c(\mbox{\boldmath$\alpha$}^{-}) \cup c(\mbox{\boldmath$\alpha$}^{+}_J)$. 
We define sequence $({b_k^{'}})_{2sm-t}$ by 
a subsequence of $(b_k)_{2sm}$ such that ${b_k^{'}} > M$ for 
$k=1, 2, \ldots, 2sm-t$. We denote it 
as a set by $\mbox{\boldmath$b$}^{'}_{2sm-t}$.  Here we note 
$\mbox{\boldmath$b$}^{'}_{2sm-t}=c(\mbox{\boldmath$\alpha$}_K^{+})$.  

We define sets $Z$ and $K$ by 
$Z=\Sigma_M \setminus \widetilde{\mbox{\boldmath$b$}_t}$ 
and $K=\Sigma_{2sm} \setminus {\mbox{\boldmath$b$}^{'}_{2sm-t}}$, 
respectively.   
We define a sequence $(z(\alpha))_{M-t}$ by putting the elements of 
$Z$ in increasing order: $z(1) < z(2) < \cdots < z(M-t)$ where  
$Z=\{ z(i); i=1, 2, \ldots, M-t \}$, 
and a sequence $(\kappa_j)_t$     
 by putting the elements of $K$ in increasing order: 
$\kappa_1 < \kappa_2 < \cdots < \kappa_{t}$ where  
$K=\{ \kappa_j; \, j=1, 2, \ldots, t \}$.  

We derive the spin-$s$ correlation functions from 
those of the spin-1/2 case sending $\epsilon$ to zero:  
\begin{equation} 
F_{m}^{(2s \, +)}(\{ i_b, j_b \}; (w_j^{(2s; \, +)})_{L} )= 
C(\{ i_k, j_k \}) \, 
\lim_{\epsilon \rightarrow 0} 
F_{2sm}^{(1 \, +)}(\{\epsilon_j, \epsilon_j^{'} \} ; 
(w_j^{(2s; \, \epsilon)})_{L} ) \, . 
\end{equation}
Applying (\ref{eq:FCR}) to (\ref{eq:def-CF}) (or (\ref{eq:CFe})) we have 
\begin{eqnarray} 
& &  
F_{2sm}^{(1 \, +)}(\{\epsilon_j, \epsilon_j^{'} \};  
(w_j^{(2s; \, \epsilon)})_{L}) 
= \sum_{t=r}^{2sm} \sum_{\{ c_j, c_j^{'} \}_t}
G_{\{ c_j, c_j^{'} \}}(\lambda_1, \cdots, \lambda_{M+2sm}) 
 \nonumber \\ 
&  & \times \phi_{2sm}(\{ \lambda_{\alpha} \}_M) 
\frac 
{\langle 0 | \prod_{\alpha=1}^{M-t} C(\lambda_{z(\alpha)}) 
             \prod_{\gamma=1}^{t} C(w_{\kappa_{\gamma}}) 
             \prod_{\beta=1}^{M-t} B(\lambda_{z(\beta)}) 
             \prod_{\gamma=1}^{t} B(\lambda_{\widetilde{b_{\gamma}}}) 
 | 0 \rangle} 
{\langle 0 | \prod_{\alpha=1}^{M-t} C(\lambda_{z(\alpha)}) 
             \prod_{\gamma=1}^{t} C(w_{\widetilde{b_{\gamma}}}) 
             \prod_{\beta=1}^{M-t} B(\lambda_{z(\beta)}) 
             \prod_{\gamma=1}^{t} B(\lambda_{\widetilde{b_{\gamma}}}) 
 | 0 \rangle}  \nonumber \\   
& = & 
\sum_{t=r}^{2sm} \sum_{\{ c_j, c_j^{'} \}_t} 
\prod_{\alpha=1}^{M} \prod_{j=1}^{2sm} 
        {\frac {\varphi(\lambda_{\alpha}- w_j)}  
               {\varphi(\lambda_{\alpha}- w_j +\eta)}}  
\prod_{j \in \mbox{\boldmath$\alpha$}^{-}} 
    \left( \frac {\prod_{b=1, b \in A_j}^{M+j-1} 
                  \varphi(\lambda_{c_j}-\lambda_b+\eta)}   
             {\prod_{b=1, b \in A_j^{'}}^{M+j} 
              \varphi(\lambda_{c_j}-\lambda_b)}   
\right) 
\nonumber  \\ 
& &  \times 
\prod_{j \in \mbox{\boldmath$\alpha$}^{+}} 
\left( \frac {\prod_{b=1, b \in A_j^{'}}^{M+j-1} 
              \varphi(\lambda_{b}-\lambda_{c_j^{'}}+\eta)}   
             {\prod_{b=1, b \in A_{j+1}}^{M+j} 
              \varphi(\lambda_{b}-\lambda_{c_j^{'}})}   
\right) 
\prod_{1 \le k < \ell \le t} 
{\frac {\varphi(\lambda_{\widetilde{b_k}}-\lambda_{\widetilde{b_{\ell}}})}
       {\varphi(w_{\kappa_k}-w_{\kappa_{\ell}})} }
\nonumber  
\end{eqnarray} 
\begin{eqnarray} 
& & \times \, \prod_{\alpha=1}^{M-t} \prod_{\ell=1}^{t} 
\frac {\varphi(\lambda_{z(\alpha)}-\lambda_{\widetilde{b_{\ell}}})}  
      {\varphi(\lambda_{z(\alpha)}-w_{\kappa_{\ell}})}  
\prod_{\alpha=1}^{M} \prod_{\ell=1}^{t} 
\frac {\varphi(\lambda_{\alpha}-w_{\kappa_{\ell}}+\eta)}  
      {\varphi(\lambda_{\alpha}-\lambda_{\widetilde{b_{\ell}}}+\eta)}  
\nonumber \\ 
& &  \quad \times \, \, 
\det \Bigg( (\Phi^{'})^{-1}\left( \lambda_{z(\alpha)})_{M-t} \# 
(\lambda_{\widetilde{b_{\ell}}} )_t \right) \times \nonumber \\   
& & \qquad \qquad \qquad \times 
\Psi^{'} \left( (\lambda_{z(\alpha)})_{M-t} \# (w_{\kappa_{\ell}})_t, 
\,   (\lambda_{z(\alpha)})_{M-t} \# 
     (\lambda_{\widetilde{b_{\ell}}} )_t \right) \Bigg) \, . 
\label{eq:ratios} 
\end{eqnarray} 
Here, $\phi_{m}(\{ \lambda_{\alpha} \}) = \prod_{j=1}^{m} 
\prod_{\alpha=1}^{M} b(\lambda_{\alpha}-w_j)$, and   
matrix elements $(\Psi^{'})_{ab}$ for $a, b=1, 2, \ldots, M$ are given by 
\begin{eqnarray} 
& & \left(\Psi^{'}( (\lambda_{z(\alpha)})_{M-t} \# (\lambda_{\widetilde{b_{\ell}}} )_t , \, (\lambda_{z(\alpha)})_{M-t} \# (w_{\kappa_{\ell}})_t; 
(w_k)_{L} ) \right)_{a, \,  b} 
\nonumber \\ 
& &  = \left\{ 
\begin{array}{cc}  
\Phi^{'}_{a, \, b} \left( (\lambda_{z(\alpha)})_{M-t} \# 
(\lambda_{\widetilde{b_{\ell}}} )_t \right) 
 &  \, \, \mbox{for} \, \,   b \le M-t  \\ 
\displaystyle{\frac {\varphi(\eta)}
{\varphi(\lambda_{z(a)} - w_{\kappa_k}) 
\varphi(\lambda_{z(a)} - w_{\kappa_k}+\eta)}}
 &   \, \,  \mbox{for} \,  b=k+M-t \, \,  (1 \le k \le t)     
\end{array}
\right.   
\end{eqnarray}
The matrix elements of the Gaudin matrix are given as follows. 
\begin{eqnarray}
& & \Phi^{'}_{a, \, b}\left( (\lambda_{z(\alpha)})_{M-t} \# 
(\lambda_{\widetilde{b_{\ell}}} )_t ; (w_k)_L \right) = 
\Phi^{'}_{z(a), z(b)}((\lambda_{\alpha})_{M} ; (w_k)_L  ) 
\nonumber \\ 
& & = {\frac {\varphi(2 \eta)} 
      {\varphi(\lambda_a-\lambda_b+ \eta) 
       \varphi(\lambda_a- \lambda_b - \eta)}} 
+ \delta_{a, b} \Bigg( \sum_{p=1}^{L} 
{\frac {\varphi(\eta)} 
      {\varphi(\lambda_a- w_p) \varphi(\lambda_a- w_p + \eta)}}
\nonumber \\ 
& & \qquad 
- \sum_{\gamma=1}^{M} \frac {\varphi(2 \eta)} 
{\varphi(\lambda_a - \lambda_{\gamma} + \eta)
 \varphi(\lambda_a - \lambda_{\gamma} - \eta)} \Bigg) \, . 
\end{eqnarray}

For any positive integer $N_s$ we can rigorously calculate (\ref{eq:ratios}) 
as follows. \cite{Deg} 
\begin{proposition} 
\begin{eqnarray} 
& & F_{2sm}^{(1 \, +)}(\{ \epsilon_j, \epsilon_j^{'} \}; 
(w_j^{(2s; \, \epsilon)})_L )  
 = \sum_{t=r}^{2sm} \sum_{\{c_j, c_j^{'} \}} \, 
\left( \prod_{j, k \in \mbox{\boldmath$\alpha$}^{+}_K,
 c_j^{'} < c_k^{'}, j < k } (-1) \right) 
\nonumber \\  
& & 
\times (-1)^{2sm-t} \prod_{j \in \mbox{\boldmath$\alpha$}^{+}_K} 
\left( \prod_{\ell \in \mbox{\boldmath$\alpha$}^{+}_J; \ell> j} (-1) \cdot 
 \prod_{\kappa \in K; \kappa+M < c_j^{'} } (-1) \right)  
\nonumber \\ 
%
& & \times 
\det (\Phi^{'})^{-1} \Psi^{'}( (\lambda_{z({\alpha})})_{M-t} \# 
(\xi_{\kappa_{\ell}})_t, \, 
(\lambda_{z({\alpha})})_{M-t} \# (\lambda_{\widetilde{b_{\ell}}})_t) \times  
\nonumber \\ 
& & 
 \times \, (-1)^{r^{'}} \,   
{\frac { \prod_{j \in \mbox{\boldmath$\alpha$}^{-}} \left( \prod_{k=1}^{j-1} 
\varphi(\lambda_{c_j} - w_k^{(2s; \, \epsilon)} + \eta) 
\prod_{k=j+1}^{2sm} \varphi(\lambda_{c_j} - w_k^{(2s; \, \epsilon)} ) \right)}
{\prod_{1 \le k < \ell \le 2sm} 
\varphi(\lambda_{b_{\ell}} - \lambda_{b_k} + \eta)} } \nonumber 
\end{eqnarray}
\begin{eqnarray}
& & \times  
{\frac { \prod_{j \in \mbox{\boldmath$\alpha$}^{+}} 
\left( \prod_{k=1}^{j-1} 
\varphi(\lambda_{c_j^{'}} - w_k^{(2s; \, \epsilon)} - \eta) 
\prod_{k=j+1}^{2sm} 
\varphi(\lambda_{c_j^{'}} - w_k^{(2s; \, \epsilon)} ) \right)}
{\prod_{1 \le k < \ell \le 2sm} 
\varphi(w_{k}^{(2s; \, \epsilon)} - w_{\ell}^{(2s; \, \epsilon)})} } \, . 
\nonumber \\ 
\label{eq:finite-sum}
\end{eqnarray} 
\end{proposition} 
We define matrix elements $(j, k)$ of $\phi_M^{(2sm)}((b_{\ell})_{2sm})$ 
($1 \le j \le 2sm$): \cite{Deg}  
\begin{eqnarray} 
& & \mbox{If} \quad  b_{j} > M,  \quad 
\left( \phi_M^{(2sm)}((b_{\ell})_{2sm}) \right)_{j, \,  k} 
= - \delta_{b_{j}-M, k} \quad 
\mbox{for} \, k=1, 2, \ldots, 2s m,  \nonumber \\ 
& & \mbox{if} \quad  b_{j} \le M,  \quad 
\mbox{there is an integer} \, i \, \mbox{such that} 
\, b_j= {\tilde b}_{i} \,   \nonumber \\ 
& & \quad \left( \phi_M^{(2sm)}((b_{\ell})_{2sm}) \right)_{j, \,  \kappa_k} 
= (\Phi^{'})^{-1} \Psi^{'}( (\lambda_{z({\alpha})})_{M-t} \# 
(\lambda_{\widetilde{b_{\ell}}})_t, \, 
\nonumber \\ 
& & \qquad \qquad \qquad 
(\lambda_{z({\alpha})})_{M-t} \# (\xi_{\kappa_{\ell}})_t 
 )_{i+M-t, k+M-t} \, , \quad \mbox{for} \, k=1, 2, \ldots, t,
\nonumber \\ 
& & \quad \mbox{and}  
\quad  \phi_M^{(2sm)}((b_{\ell})_{2sm})_{j, b_k^{'}} = 0 \quad 
\mbox{for} \, k=1, 2, \ldots, 2sm - t. 
\end{eqnarray}
We can show the following proposition. \cite{Deg}
\begin{proposition} 
\begin{eqnarray} 
& & \det \left( (\Phi^{'})^{-1} \Psi^{'}( (\lambda_{z({\alpha})})_{M-t} \# 
(w_{\kappa_{\ell}})_t, \, 
(\lambda_{z({\alpha})})_{M-t} \# (\lambda_{\widetilde{b_{\ell}}})_t) 
\right)   \nonumber \\ 
& = & \det \phi_M^{(2sm)}((b_{\ell})_{2sm})  \,  (-1)^{2sm-t} \, 
\left( \prod_{j, k \in \mbox{\boldmath$\alpha$}^{+}_K,
 c_j^{'} < c_k^{'}, j < k } (-1) \right) 
 \nonumber \\ 
& & \qquad \times 
\prod_{j \in \mbox{\boldmath$\alpha$}^{+}_K} 
\left( \prod_{\ell \in \mbox{\boldmath$\alpha$}^{+}_J; \ell> j} (-1) \cdot 
 \prod_{\kappa \in K; \kappa+M < c_j^{'} } (-1) \right) \, . 
\label{eq:Phi-Psi} 
\end{eqnarray} 
\end{proposition}

When $N_s$ is large enough,  
solving the integral equations for $\phi_M^{(2sm)}((b_{\ell})_{2sm})$, 
we can show 
\begin{equation} 
\det \phi_M^{(2sm)}((b_{\ell})_{2sm})= \det M^{(2sm)}((b_{\ell})_{2sm}) 
+ O(1/N_s). 
\end{equation} 
We thus obtain  
the finite-size spin-$s$ XXZ correlation functions with arbitrary entries 
(\ref{eq:finiteCF}).

\section*{Acknowledgments}
One of the authors (T.D.) would also like to thank K. Motegi for 
helpful collaboration on the spin-$s$ $R$-matrices in Ref. 
\refcite{DMo}, which are closely related to the present study.    
Furthermore, the authors would like to thank 
S. Miyashita for encouragement and keen interest in this work.  
The authors are grateful to the organizers of the workshop 
``Infinite Analysis 09-- New Trends in Quantum Integrable Systems --'', 
July 27-31, 2009, 
Kyoto University, Japan.   
This work is partially supported by 
Grant-in-Aid for Scientific Research (C) No. 20540365.

\begin{verbatim}
\bibliographystyle{ws-procs9x6}
%\bibliography{Deguchi}
\end{verbatim}

\end{document}